\documentclass[applsci,article,publish,moreauthors]{mdpi}
\UseRawInputEncoding
\usepackage{amsmath}
\usepackage{amsfonts}
\usepackage{amssymb,bm}
\usepackage{siunitx}
\usepackage{color}
\usepackage{braket}
\usepackage{graphics}
\firstpage{1}
\pubvolume{13}
\issuenum{13}
\articlenumber{7511}
\pubyear{2023}
\copyrightyear{2023}
\externaleditor{Academic Editor: Petr Korusenko}
\datereceived{26 May 2023 }
\dateaccepted{23 June 2023 }
\datepublished{25 June 2023 }
\hreflink{https://doi.org/10.3390/app13137511} % If needed use \linebreak
\Title{Impact of Mass-Gap on the Dispersion Interaction of Nanoparticles with
Graphene out of Thermal Equilibrium}

\TitleCitation{Impact of Mass-Gap on the Dispersion Interaction of Nanoparticles with
Graphene out of Thermal Equilibrium}

\Author{{Galina L. Klimchitskaya} ${}^{1,2}$*\orcidA{},
Constantine C. Korikov${}^{3}$,
Vladimir M. Mostepanenko ${}^{1,2,4,}$\orcidB{} and
Oleg {Yu.} Tsybin ${}^{2}$}

\AuthorNames{Galina L. Klimchitskaya, Constantine C. Korikov,
Vladimir M. Mostepanenko, Oleg Yu. Tsybin}

\AuthorCitation{Klimchitskaya, G.L.; Korikov, C.C.; Mostepanenko, V.M.; Tsybin, O.Y.}
\address{%
$^{1}$ \quad Central Astronomical Observatory at Pulkovo of the Russian Academy of Sciences, 196140 Saint Petersburg, Russia\\
$^{2}$ \quad Peter the Great Saint Petersburg
Polytechnic University, 195251 Saint Petersburg, Russia\\
$^{3}$ \quad Huawei Noah's Ark Lab, Krylatskaya str. 17, Moscow 121614, Russia \\
$^{4}$ \quad {Kazan Federal University}, 420008 Kazan, Russia}

\corres{Correspondence: g.klimchitskaya@gmail.com}

\abstract{We consider the nonequilibrium dispersion force acting on
nanoparticles on the source side of gapped graphene sheet. Nanoparticles
are kept at the environmental temperature, whereas the graphene sheet may
be either cooler or hotter than the environment. Calculation of the
dispersion force as a function of separation at different values of the
mass-gap parameter is performed using the generalization of the
fundamental Lifshitz theory to the out-of-thermal-equilibrium conditions.
The response of gapped graphene to quantum and thermal fluctuations of
the electromagnetic field is described by the polarization tensor in
(2+1)-dimensional space-time in the framework of the Dirac model. The
explicit expressions for the components of this tensor in the area of
evanescent waves are presented. The nontrivial impact of the mass-gap
parameter of graphene on the nonequilibrium dispersion force, as
compared to the equilibrium one, is determined. It is shown that,
unlike the case of a pristine graphene, the nonequilibrium force
preserves an attractive character. The possibilities of using the
obtained results in the design of micro- and nanodevices incorporating
nanoparticles and graphene sheets for their functionality are discussed.}

\keyword{dispersion force; thermal nonequilibrium; nanoparticles;
Lifshitz theory; graphene; polarization tensor; nanodevices}

\begin{document}

%%%%%%%%%%%%%%%%%%%%%%%%%%%%%%%%%%%%%%%%%%

\section{Introduction}

Investigation of interaction between nanoparticles and material surfaces
of different nature is of profound importance for physics and its
applications in nanotechnology, including bioelectronics (see, e.g., the
articles and reviews \cite{1,2,3,4,5,6,7,8,9,10,11,12,13,14,15,15a,15b}). The
microparticle-surface interaction includes several contributions, among
which are mechanical contact forces, Born repulsion, and attractive
dispersion forces \cite{16,17}. At separations between a nanoparticle and
a surface exceeding several nanometers, the dispersion forces, which are
also called the van der Waals or Casimir-Polder forces, become dominant.
They are determined by the quantum and thermal fluctuations of the
electromagnetic field.

The entirely new material, which finds increasing use in nanotechnology,
is graphene, i.e., the plane sheet of carbon atoms arranged in a
hexagonal lattice \cite{18,19,20}. The dispersion (Casimir-Polder)
interaction of graphene with different atomic systems
\cite{21,22,23,24,25,26,27,28,29,30,31,32,33,34} and nanoparticles
\cite{35,36,37,38,39,40} has been the subject of much investigation. The
obtained results are finding ever-widening application in bioelectronics
\cite{41,42,43,44}.

The implementation of interaction between nanoparticles and graphene to
new generation of nanodevices called for a development of theoretical
methods which make it possible to calculate the dispersion force as a
function of all relevant parameters. These methods have been developed
in the framework of the Lifshitz theory \cite{45,46,47}, by expressing
the dispersion force between an atom or a nanoparticle and a graphene
sheet via the atomic (nanoparticle) electric polarizability and the
polarization tensor of graphene
\cite{21,22,23,24,25,26,27,28,29,30,31,32,33,34}. In so doing, the
polarization tensor of graphene was found \cite{48,49,50,51} on the
basis of first principles of thermal quantum field theory in the
framework of the Dirac model \cite{18,19,20}.

The Lifshitz theory of dispersion forces is formulated for the case
when the interacting bodies are in the state of thermal equilibrium
with the environment. This condition, however, is violated when both
of the interacting bodies (or at least one of them) are kept at
temperatures different from that of the environment. The formalism
generalizing the Lifshitz theory for systems out of thermal equilibrium
was developed in \cite{52,53,54,55,56,57}. During the last few years,
different aspects of the nonequilibrium dispersion forces acting
between two material plates, a small sphere or an atom and a material
plate and between two spheres were investigated using this formalism
\cite{58,59,60,61,62,63,64}. Specifically, the case of
temperature-dependent response functions of the interacting bodies
was considered in \cite{61,62}.

The nonequilibrium dispersion force acting on spherical nanoparticles
on the source side of an ideal (pristine) freestanding in vacuum
graphene sheet was investigated quite recently \cite{65}. The pristine
character of graphene assumed in \cite{65} means that its crystal
lattice does not include any foreign atoms and the quasiparticles are
massless, as was supposed in the original Dirac model \cite{18,19,20}.
Reference \cite{65} suggested that the temperature of nanoparticles is
the same as of the environment, whereas the graphene sheet can be
either cooler or hotter than the environment. It was shown that an
impact of the nonequilibrium effects of the dispersion force decreases
with increasing graphene-nanoparticle separation distance. What is more,
according to the results obtained, at relatively short separations the
effects of nonequilibrium may change the sign of the dispersion force
by making it repulsive \cite{65}.

In this article, we apply the theory of nonequilibrium dispersion
interaction to investigate the force acting on nanoparticles kept at the
environmental temperature on the source side of gapped graphene described
by the Dirac model with light but massive quasiparticles. The temperature
of a graphene sheet is assumed to be either lower or higher than that of
the environment. To perform computations of the dispersion force in this
case, we present the explicit expressions for the polarization tensor
of gapped graphene along the real frequency axis in the region of
evanescent waves, which have not been considered in the literature up to
now with sufficient detail. We demonstrated that the value of the mass-gap
parameter makes a nontrivial impact on the nonequilibrium force, as
compared to the equilibrium one, depending on the values of separation
and graphene temperature. Unlike the case of a pristine graphene, for a
nonzero mass of quasiparticles the nonequlibrium dispersion force
preserves its attractive character.

The structure of the article is as follows. In Section 2, we present the
expression for a nonequilibrium dispersion force acting on nanoparticles
on the source side of gapped graphene sheet in terms of the polarization
tensor. In Section 3, the components of this tensor in the area of the
evanescent waves are specified. Section 4 contains the computational
results for the dispersion force acting on nanoparticles which is shown
as the function of separation for different values of the mass-gap
parameter and at different temperatures. In Sections 5 and 6, the reader
will find the discussion of the obtained results and our conclusions.

%%%%%%%%%%%%%%%%%%%%%%%%%%%%%%%%%%%%%%%%%%
\section{Nonequilibrium Dispersion Force on a Nanoparticle on the
Source Side of Gapped Graphene }
%%%%%%%%%%%%%%%%%%%%%%%%%%%%%
\newcommand{\adt}{(a,\Delta,T_E,T_g)}
\newcommand{\ok}{(\omega,k;\Delta,T_g)}
\newcommand{\wk}{(\omega,k)}
\newcommand{\fok}{(\omega,k;\Delta)}
\newcommand{\sok}{(\omega,k,\Delta)}
\newcommand{\tp}{\tilde{p}}
%%%%%%%%%%%%%%%%%%%%%%%%%%%%

We consider the dispersion (Casimir-Polder) force acting on a spherical nanoparticle of radius $R$
spaced above a graphene sheet at a separation $a\gg R$. The consideration of nanoparticles of
other types (for instance, having a nonspherical shape) would need a more complicated
theory using the scattering-matrix approach \cite{56}. The area of graphene sheet is taken to
be much larger than the separation to a nanoparticle squared. It is assumed that at all
temperatures $T$ under consideration it holds $R\ll\hbar c/(k_BT)$, where $k_B$ is the Boltzmann constant
[for instance, at the environmental temperature $T_E=300~$K one has
$\hbar c/(k_BT)\approx 7.6~\mu$m]. Under this condition, within the range of separations $a$
considered below, the nanoparticle can be described by the static polarizability $\alpha(0)$,
which takes the form \cite{59}

\begin{equation}
\alpha(0)=R^3\frac{\varepsilon(0)-1}{\varepsilon(0)+1}, \qquad
\alpha(0)=R^3
\label{eq1}
\end{equation}

\noindent
for dielectric and metallic nanoparticles, respectively, where $\varepsilon(0)$ is the static
dielectric permittivity of a nanoparticle material.

Below we assume that nanoparticles have the same temperature $T_E$ as the environment,
whereas the graphene sheet has the temperature $T_g$ which is either lower or higher than $T_E$.
As distinct from \cite{65}, where the case of a pristine graphene was considered, here the
graphene sheet is characterized by a nonzero mass-gap parameter $\Delta=2mv_F^2$, where $m$
is the mass of quasiparticles and $v_F\approx c/300$ is the Fermi velocity \cite{19,66,67}.

The nonequilibrium dispersion force, acting on a nanoparticle, is represented in the form
\cite{54,56}

\begin{equation}
F_{\rm neq}\adt =F_{M}\adt+F_r\adt,
\label{eq2}
\end{equation}

\noindent
where $F_{\rm M}\adt$ can be expressed as a sum over the discrete Matsubara frequencies,
much as the equilibrium Casimir-Polder force \cite{68,69}, whereas $F_r\adt$  is the
contribution  which is given by an integral over the real frequency axis.

In fact, the effects of nonequilibrium contribute to both terms in the right-hand side of
(\ref{eq2}). Because of this, it is not reasonable to call the first of them "equilibrium"
and the second --- "nonequilibrium" that occurs in the literature. Moreover, the division
of $F_{\rm neq}$ into $F_{\rm M}$ and $F_r$ is not unique and can be made in a number of
ways. Below we use the same division as in \cite{65}.

In this case, the first term in (\ref{eq2}) is given by \cite{65}

\begin{eqnarray}
&&
F_{M}\adt=-\frac{2k_BT_E\alpha(0)}{c^2}\sum_{l=0}^{\infty}{\vphantom{\sum}}^{\prime}
\int\limits_0^{\infty}k\,dke^{-2aq_l(k)}
\label{eq3} \\
&&~~~~~\times
\left\{\left[2q_l^2(k)c^2-\xi_{E,l}^2\right]
R_{\rm TM}(i\xi_{E,l},k;\Delta,T_g)-\xi_{E,l}^2R_{\rm TE}(i\xi_{E,l},k;\Delta,T_g)\right\}.
\nonumber
\end{eqnarray}

\noindent
Here, $k$ is the magnitude of the wave vector component along the graphene sheet,
$q_l^2(k)=k^2+\xi_{E,l}^2/c^2$, $\xi_{E,l}=2\pi k_BT_El/\hbar$ with $l=0,\,1,\,2,\,\ldots$
are the Matsubara frequencies at the environmental temperature $T_E$, and the prime on the
summation sign multiples the term with $l=0$  by the factor 1/2.

The quantities $R_{\rm TM}$ and $R_{\rm TE}$ are the reflection coefficients of
the electromagnetic fluctuations on a graphene sheet for the transverse magnetic (TM) and
transverse electric (TE) polarizations calculated at the pure imaginary Matsubara
frequencies $\omega=i\xi_{E,l}$, but at the temperature of graphene $T_g$.  They are
expressed via the components of the polarization tensor of graphene $\Pi_{ij}\ok$
\cite{49,50,70}

\begin{eqnarray}
&&
R_{\rm TM}\ok=\frac{q(\omega,k)\Pi_{00}\ok}{2\hbar k^2+q(\omega,k)\Pi_{00}\ok},
\nonumber \\
&&
R_{\rm TE}\ok=-\frac{\Pi\ok}{2\hbar k^2q(\omega,k)+\Pi\ok},
\label{eq4}
\end{eqnarray}

\noindent
where $q^2(\omega,k)=k^2-\omega^2/c^2$ and the quantity $\Pi$ is defined as

\begin{equation}
\Pi\ok=k^2\Pi_i^{\,i}\ok-q^2(\omega,k)\Pi_{00}\ok
\label{eq5}
\end{equation}

\noindent
with the summation over the repeated index $i=0,\,1,\,2$. The explicit expressions for the
polarization tensor in the required frequency regions are given in the next section.

Note that the polarization tensor $\Pi_{ij}$ describing
the response of graphene to quantum and thermal fluctuations of the electromagnetic field
strongly depends on temperature $T_g$ as a parameter. In application to the nonequilibrium
dispersion forces, similar situation was considered previously  for the phase-change \cite{61}
and metallic \cite{62} materials.

The second term on the right-hand side of (\ref{eq2}), according to the division accepted in
\cite{65}, takes the form \cite{56}

\begin{eqnarray}
&&
F_{r}(a,\Delta, T_E,T_g)=\frac{2\hbar\alpha(0)}{\pi c^2}
\int\limits_{0}^{\infty}\!\!d\omega\Theta(\omega,T_E,T_g)\!\!
\int\limits_{\omega/c}^{\infty}\!\!k\,dke^{-2aq(\omega,k)}
\nonumber \\
&&~~~~\times
{\rm Im}\left\{\left[2q^2(\omega,k)c^2+\omega^2\right]
R_{\rm TM}\ok+\omega^2R_{\rm TE}\ok\right\}.
\label{eq6}
\end{eqnarray}

\noindent
Here, the quantity $\Theta(\omega,T_E,T_g)$ is defined as
\begin{equation}
\Theta(\omega,T_E,T_g)=\frac{1}{\exp\left(\frac{\hbar\omega}{k_BT_E}\right)-1}-
\frac{1}{\exp\left(\frac{\hbar\omega}{k_BT_g}\right)-1}.
\label{eq7}
\end{equation}

The important property of the division (\ref{eq2}) accepted in \cite{65} is that the
quantity $F_r$ expressed in terms of real frequencies is determined by the
contribution of only the evanescent waves for which $k>\omega/c$. As a result, the exponent
in (\ref{eq6}) has the real power. This is advantageous as compared to the standard
Lifshitz formula for equilibrium Casimir and Casimir-Polder forces written in terms of
real frequencies, which contains the contributions of both the evanescent and propagating
($k<\omega/c$) waves \cite{69}. For the latter contribution, the quantity $q(\omega,k)$
is pure imaginary resulting in the integral of quickly oscillating function which makes
integration difficult.

%%%%%%%%%%%%%%%%%%%%%%%%%%%%%%%%%%%%%%%%%%
\section{Polarization Tensor in the Area of Evanescent Waves}

The nonequilibrium dispersion force (\ref{eq2}) acting on nanoparticles on the source of a
 gapped graphene sheet can be computed by Equations (\ref{eq3})--(\ref{eq7}).
 For this purpose, one should know the component of the polarization tensor $\Pi_{00}$
 and the combination of its components $\Pi$ defined in (\ref{eq5}) for a graphene sheet
 with the nonzero mass-gap parameter $\Delta$. As mentioned in Section 1,
the polarization tensor of graphene was found in \cite{48,49,50,51} in the framework
of the Dirac model. In doing so, \cite{48} was devoted to the case of zero temperature,
$T=0$. In \cite{49} the polarization tensor of graphene was obtained at nonzero
temperature at all discrete Matsubara frequencies. These results, however, did not admit
a continuation to the entire plane of complex frequencies and, specifically, were
inapplicable along the real frequency axis. Thus, they can be used for calculation of
the equilibrium Casimir and Casimir-Polder forces and the contribution $F_M$ to the
nonequilibrium force, but not the contribution $F_r$.

The polarization tensor of graphene with nonzero $\Delta$ valid over the entire plane of
complex frequencies was derived in \cite{50}, where the most attention was paid to the
region of propagating waves $k<\omega/c$ in connection with the topical applications to
the reflectivity \cite{71,72,73,74} and conductivity \cite{75,76,77,78} properties of
graphene. Below we present a more detailed exposition of the results of \cite{50} relevant
to the area of evanescent waves ($k>\omega/c$) which determine the contribution (\ref{eq6})
to the nonequilibrium dispersion force.

Before dealing with the polarization tensor, a few remarks concerning the area of application
of this quantity for the calculation of dispersion forces are in order. In \cite{48,49,50,51},
the polarization tensor of graphene was derived in the framework of the Dirac model.
This model provides the physically adequate description of graphene at energies below
approximately 3~eV \cite{79}. Thus, the energies $\hbar\omega$ giving the major contribution
to the dispersion force should be below this limit. The characteristic frequency determining
the dispersion force is $\omega_c=c/(2a)$ \cite{68,69}. It is easily seen that the respective
characteristic energy $\hbar\omega_c$ is below 1~eV at all separations $a>100~$nm.
Therefore, at separations, say, $a>200~$nm one can safely use the Dirac model and its
consequences in calculations of dispersion forces. This was confirmed by the fact that
measurements of the dispersion interaction with graphene were found in a very good
agreement with theoretical predictions computed using the polarization tensor \cite{80,81}.

Now we present the explicit expressions for the quantities $\Pi_{00}$ and $\Pi$ in the
frequency region of evanescent waves $\omega/c<k$. Similar to \cite{28}, we present these
quantities as the sums of two contributions

\begin{eqnarray}
&&
\Pi_{00}\ok=\Pi_{00}^{(0)}\fok+\Pi_{00}^{(1)}\ok,
\nonumber \\
&&
\Pi\ok=\Pi^{(0)}\fok+\Pi^{(1)}\ok.
\label{eq8}
\end{eqnarray}

\noindent
Here, the contributions with an upper index (0) are defined at zero temperature, $T=0$,
whereas the quantities with an upper index (1) have a meaning of the thermal corrections
to them. In doing so, both contributions depend on the mass-gap parameter of graphene
$\Delta$. With vanishing temperature, both $\Pi_{00}^{(1)}$ and $\Pi^{(1)}$ go to zero.

The analytic continuation of the polarization tensor of graphene to the frequency region
of evanescent waves takes different forms in the  interval

\begin{equation}
\frac{\omega}{c}<k\leqslant\frac{\omega}{v_F}\approx 300\frac{\omega}{c}
\label{eq9}
\end{equation}

\noindent
and in the interval

\begin{equation}
300\frac{\omega}{c}\approx\frac{\omega}{v_F} <k<\infty.
\label{eq10}
\end{equation}

First, we consider the interval (\ref{eq9}) which is often called the plasmonic region
\cite{82}. In this region the first contributions to (\ref{eq8}) take the form \cite{50}

\begin{eqnarray}
&&
\Pi_{00}^{(0)}\fok=-\frac{2\alpha k^2}{cp^2\wk}\,\Phi\sok,
\nonumber \\
&&
\Pi^{(0)}\fok=\frac{2\alpha k^2}{c}\Phi\sok,
\label{eq11}
\end{eqnarray}

\noindent
where

\begin{equation}
p^2\wk=\frac{\omega^2}{c^2}-\frac{v_F^2}{c^2}k^2\geqslant 0
\label{eq13}
\end{equation}

\noindent
and $\alpha=e^2/(\hbar c)$ is the fine structure constant.
The function $\Phi$ is defined as

\begin{equation}
\Phi\sok=\Delta-\hbar cp\wk\left[1+\frac{\Delta^2}{\hbar^2c^2p^2\wk}\right]\,
\left[{\rm arctanh} \frac{\Delta}{\hbar c p\wk}+i\frac{\pi}{2}\right]
\label{eq14}
\end{equation}

\noindent
for $\hbar c p\wk\geqslant\Delta$ and as

\begin{equation}
\Phi\sok=\Delta-\hbar cp\wk\left[1+\frac{\Delta^2}{\hbar^2c^2p^2\wk}\right]\,
{\rm arctanh} \frac{\hbar c p\wk}{\Delta}
\label{eq15}
\end{equation}

\noindent
for $\hbar c p\wk<\Delta$.

The second contributions to (\ref{eq8}) in the plasmonic region are more complicated.
It is convenient to define their real and imaginary parts separately. We start from
defining the real parts of $\Pi_{00}^{(1)}$ and $\Pi^{(1)}$ which, in turn, have different
forms under the conditions $\hbar c p\wk\geqslant\Delta$ and $\hbar c p\wk<\Delta$.

Thus, if the condition $\hbar c p\wk\geqslant\Delta$ is satisfied, one obtains from
\cite{50} after identical transformations

\begin{equation}
{\rm Re}\,\Pi_{00}^{(1)}\ok=\frac{8\alpha\hbar c^2}{v_F^2}(I_1+I_2+I_3),
\label{eq16}
\end{equation}

\noindent
where the following notations are introduced:

\begin{eqnarray}
&&
I_1=\int\limits_{\frac{\Delta}{2\hbar c}}^{u^{(-)}\wk}\frac{du}{e^{\beta(T_g)u}+1}
\left[2-\frac{B_1(2cu+\omega)+B_1(2cu-\omega)}{cp\wk}\right],
\nonumber \\
&&
I_2=\int\limits_{u^{(-)}\wk }^{u^{(+)}\wk}\frac{du}{e^{\beta(T_g)u}+1}
\left[2-\frac{B_1(2cu+\omega)}{cp\wk}\right],
\label{eq17} \\
&&
I_3=\int\limits_{u^{(+)}\wk}^{\infty}\frac{du}{e^{\beta(T_g)u}+1}
\left[2-\frac{B_1(2cu+\omega)-B_1(2cu-\omega)}{cp\wk}\right].
\nonumber
\end{eqnarray}

\noindent
Here,

\begin{eqnarray}
&&
u^{(\pm)}\wk=\frac{1}{2c}[\omega\pm v_Fk\sqrt{A\fok}], \qquad
A\fok=1-\frac{\Delta^2}{\hbar^2c^2p^2\wk},
\nonumber \\
&&
B_1(x)=\frac{x^2-v_F^2k^2}{\sqrt{x^2-v_F^2k^2A\fok}}, \qquad
\beta(T_g)=\frac{\hbar c}{k_BT_g}.
\label{eq18}
\end{eqnarray}

\noindent
It is seen that all the integrals $I_j$ are the functions of $\omega$, $k$, $\Delta$,
and $T_g$.

Under the same condition $\hbar c p\wk\geqslant\Delta$, we obtain from
\cite{50}

\begin{equation}
{\rm Re}\,\Pi^{(1)}\ok=\frac{8\alpha\hbar \omega^2}{v_F^2}(J_1+J_2+J_3),
\label{eq19}
\end{equation}

\noindent
where the quantities $J_j$ are given by

\begin{eqnarray}
&&
J_1=\int\limits_{\frac{\Delta}{2\hbar c}}^{u^{(-)}\wk}\frac{du}{e^{\beta(T_g)u}+1}
\left\{2-\frac{cp\wk[B_2(2cu+\omega)+B_2(2cu-\omega)]}{\omega^2}\right\},
\nonumber \\
&&
J_2=\int\limits_{u^{(-)}\wk }^{u^{(+)}\wk}\frac{du}{e^{\beta(T_g)u}+1}
\left[2-\frac{cp\wk B_2(2cu+\omega)}{\omega^2}\right],
\label{eq20} \\
&&
J_3=\int\limits_{u^{(+)}\wk}^{\infty}\frac{du}{e^{\beta(T_g)u}+1}
\left\{2-\frac{cp\wk[B_2(2cu+\omega)-B_2(2cu-\omega)]}{\omega^2}\right\}.
\nonumber
\end{eqnarray}

\noindent
Here, the function $B_2(x)$ is defined as

\begin{equation}
B_2(x)=\frac{x^2-v_F^2k^2[1-A\fok]}{\sqrt{x^2-v_F^2k^2A\fok}}.
\label{eq21}
\end{equation}

If the opposite condition, $\hbar c p\wk<\Delta$, is satisfied, the real parts of
$\Pi_{00}^{(1)}$ and $\Pi^{(1)}$ take the form following from \cite{50}

\begin{eqnarray}
&&
{\rm Re}\,\Pi_{00}^{(1)}\ok=\frac{8\alpha\hbar c^2}{v_F^2}
\int\limits_{\frac{\Delta}{2\hbar c}}^{\infty}\frac{du}{e^{\beta(T_g)u}+1}
\nonumber \\
&&~~~\times
\left[2-\frac{B_1(2cu+\omega)-B_1(2cu-\omega)}{cp\wk}\right],
\nonumber \\[-1mm]
&&
\label{eq22} \\[1mm]
&&
{\rm Re}\,\Pi^{(1)}\ok=\frac{8\alpha\hbar \omega^2}{v_F^2}
\int\limits_{\frac{\Delta}{2\hbar c}}^{\infty}\frac{du}{e^{\beta(T_g)u}+1}
\nonumber \\
&&~~~\times
\left\{2-\frac{cp\wk[B_2(2cu+\omega)-B_2(2cu-\omega)]}{\omega^2}\right\}.
\nonumber
\end{eqnarray}

This concludes consideration of the real parts of $\Pi_{00}^{(1)}$ and $\Pi^{(1)}$
in the plasmonic region (\ref{eq9}). As to the imaginary parts of
$\Pi_{00}^{(1)}$ and $\Pi^{(1)}$, they are given by the unified expressions

\begin{eqnarray}
&&
{\rm Im}\,\Pi_{00}^{(1)}\ok=\frac{8\alpha\hbar c}{v_F^2p\wk}
\theta[\hbar c p\wk-\Delta]
\nonumber \\
&&~~\times
\int\limits_{u^{(-)}\wk}^{u^{(+)}\wk}\frac{du}{e^{\beta(T_g)u}+1}
\frac{(2cu-\omega)^2-v_F^2k^2}{\sqrt{v_F^2k^2A\fok-(2cu-\omega)^2}},
\nonumber \\
[-1mm]
&&
\label{eq23} \\[1mm]
&&
{\rm Im}\,\Pi^{(1)}\ok=\frac{8\alpha\hbar cp\wk}{v_F^2}
\theta[\hbar c p\wk-\Delta]
\nonumber \\
&&~~\times
\int\limits_{u^{(-)}\wk}^{u^{(+)}\wk}\frac{du}{e^{\beta(T_g)u}+1}
\frac{(2cu-\omega)^2+v_F^2k^2[1-A\fok]}{\sqrt{v_F^2k^2A\fok-(2cu-\omega)^2}},
\nonumber
\end{eqnarray}

\noindent
which are valid over the entire region (\ref{eq9}). Here, $\theta(x)$ is the step function
equal to 1 for $x\geqslant 0$ and to 0 for $x<0$.

Next, we consider the polarization tensor in the interval (\ref{eq10}).
In this case, the first contributions to (\ref{eq8}) are given by \cite{50}

\begin{eqnarray}
&&
\Pi_{00}^{(0)}\fok=\frac{\alpha\hbar k^2}{\tp^\wk}\,\Psi\sok,
\label{eq24} \\
&&
\Pi^{(0)}\fok={\alpha\hbar k^2}\tp\wk\,\Psi\sok,
\nonumber
\end{eqnarray}

\noindent
where

\begin{equation}
\tp^2\wk=\frac{v_F^2}{c^2}k^2-\frac{\omega^2}{c^2}\geqslant 0
\label{eq25}
\end{equation}

\noindent
and  $\Psi$ is defined as

\begin{equation}
\Psi\sok=2\left\{\frac{\Delta}{\hbar c\tp\wk}
+\left[1-\frac{\Delta^2}{\hbar^2c^2\tp^2\wk}\right]\,
\,{\rm arctan} \frac{\hbar c \tp\wk}{\Delta}\right\}.
\label{eq26}
\end{equation}

Similar to the plasmonic interval (\ref{eq9}), in the interval (\ref{eq10})
the quantities $\Pi_{00}^{(1)}$ and $\Pi^{(1)}$ are the complex-valued functions.
Here, we present their explicit expressions do not separating the real and
imaginary parts \cite{50}

\begin{eqnarray}
&&
\Pi_{00}^{(1)}\ok=\frac{8\alpha\hbar c^2\tp\wk}{v_F^2}
\int\limits_{\frac{\Delta}{\hbar c\tp\wk}}^{\infty}\frac{dv}{e^{D(\omega,k,T_g)v}+1}
\nonumber \\
&&~~~\times
\left[1-\frac{1}{2}\sum_{\lambda=\pm 1}
\frac{1-v^2-2\lambda\frac{\omega}{c\tp\wk}v}{\sqrt{1-v^2-
2\lambda\frac{\omega}{c\tp\wk}v+\frac{v_F^2k^2\Delta^2}{c^4\hbar^2\tp^4\wk}}}\right],
\nonumber \\
[-1mm]
&&
\label{eq27} \\[1mm]
&&
\Pi^{(1)}\ok=\frac{8\alpha\hbar c^2\tp\wk}{v_F^2}
\int\limits_{\frac{\Delta}{\hbar c\tp\wk}}^{\infty}\frac{dv}{e^{D(\omega,k,T_g)v}+1}
\nonumber \\
&&~~~\times
\left\{\frac{\omega^2}{c^2}-\frac{1}{2}\sum_{\lambda=\pm 1}
\frac{\left[\tp\wk v+\lambda\frac{\omega}{c}\right]^2+
\frac{v_F^2k^2\Delta^2}{c^4\hbar^2\tp^2\wk}}{\sqrt{1-v^2-
2\lambda\frac{\omega}{c\tp\wk}v+\frac{v_F^2k^2\Delta^2}{c^4\hbar^2\tp^4\wk}}}\right\},
\nonumber \\
\nonumber
\end{eqnarray}

\noindent
where

\begin{equation}
D(\omega,k,T_g)=\frac{\hbar c\tp\wk}{2k_BT_g}.
\label{eq28}
\end{equation}

For calculation of the contribution $F_M$ to the nonequilibrium dispersion force
(\ref{eq2}), which is given by (\ref{eq3}), one also needs the values of the
polarization tensor at the pure imaginary Matsubara frequencies.
They are easily obtained from (\ref{eq8}) and the respective expressions
(\ref{eq24})--(\ref{eq28}) found in the interval (\ref{eq10}) where we put
$\omega=i\xi_{E,l}$. In this case the definitions (\ref{eq25}) and (\ref{eq28})
take the form

\begin{equation}
\tp^2(i\xi_{E,l},k)\equiv\tp_l^2(k)=\frac{v_F^2}{c^2}k^2+\frac{\xi_{E,l}^2}{c^2},
\qquad
D(i\xi_{E,l},k,T_g)=\frac{\hbar c\tp_l(k)}{2k_BT_g}.
\label{eq29}
\end{equation}

Thus, all expressions for the polarization tensor appearing in both contributions
$F_M$ and $F_r$ to the nonequilibrium dispersion force through the reflection
coefficients (\ref{eq4}) are presented.

%%%%%%%%%%%%%%%%%%%%%%%%%%%%%%%%%%%%%%%%%%
\section{Computational Results for the  Dispersion Force on
Nanoparticles from Graphene}

Here, we present the computational results for the nonequilibrium
dispersion force $F_{neq}$ acting on nanoparticles of radius $R$
on the source side of a graphene sheet characterized by the mass-gap
parameter $\Delta$ which takes the typical values of 0.1 and 0.2 eV
\cite{80,81}. The temperature of nanoparticles is assumed to be the
same as of the environment, i.e., $T_E=300~$K, whereas the temperature
of a graphene sheet can be either cooler, $T_g=77~$K, or hotter,
$T_g=500~$K, than the environmental temperature. These temperatures are
chosen as a representative example. The first of them is the temperature
of liquid nitrogen, whereas the second is close to that employed in the
experiment on measuring the nonequilibrium Casimir-Polder force \cite {82a}.
The developed formalism allows computation of the nonequilibrium
dispersion interaction for any experimental temperatures.
In line with the
assumptions made in Sections 2 and 3, computations are performed in
the separation range from 200 nm to 2~$\mu$m and it is assumed that
$R$ is sufficiently small. The polarizabilities of dielectric and
metallic nanoparticles are presented in (\ref{eq1}).

Numerical computations were performed by Equations
(\ref{eq2})--(\ref{eq4}) and (\ref{eq6}), (\ref{eq7}) using the
expressions for the polarization tensor presented in Section 3. For
this purpose, we worked out a program written in the C++ programming
language. The program utilizes the Gauss-Kronrod and Double-exponential
quadrature methods from the GNU Scientific Library \cite{83} and Boost
C++ Libraries \cite{84} for numerical integration. High precision
computation is achieved with the help of the Boost Multiprecision
Library \cite{85}. The program also employs the OpenMP Library \cite{86}
for parallelization. The results presented below were obtained using
computational resources of the Supercomputer Center of the Peter the
Great Saint Petersburg Polytechnic University.

The computational results for $F_{\rm neq}$ are normalized to the classical
limit of the equilibrium dispersion force acting on a nanoparticle on
the source side of an ideal metal plane \cite{69}

\begin{equation}
F_c(a,T_E)=-\frac{3k_BT}{4a^4}\,\alpha(0).
\label{eq30}
\end{equation}

\noindent
The normalized values do not depend on the static polarizability of a
nanoparticle $\alpha(0)$. The absolute values of $F_{\rm neq}$ for the
nanoparticles made of some specific material can be obtained using
Equations (\ref{eq1}) and (\ref{eq30}) by fixing the values of $R$
and $\varepsilon(0)$. In doing so the value of $R$ is
restricted by only the conditions $R\ll a$ and $R\ll \hbar/(k_BT)$
considered in the beginning of Section 2.

\begin{figure}[H]
\vspace*{-7.2cm}
\centerline{\hspace*{-2.7cm}
\includegraphics[width=4.5in]{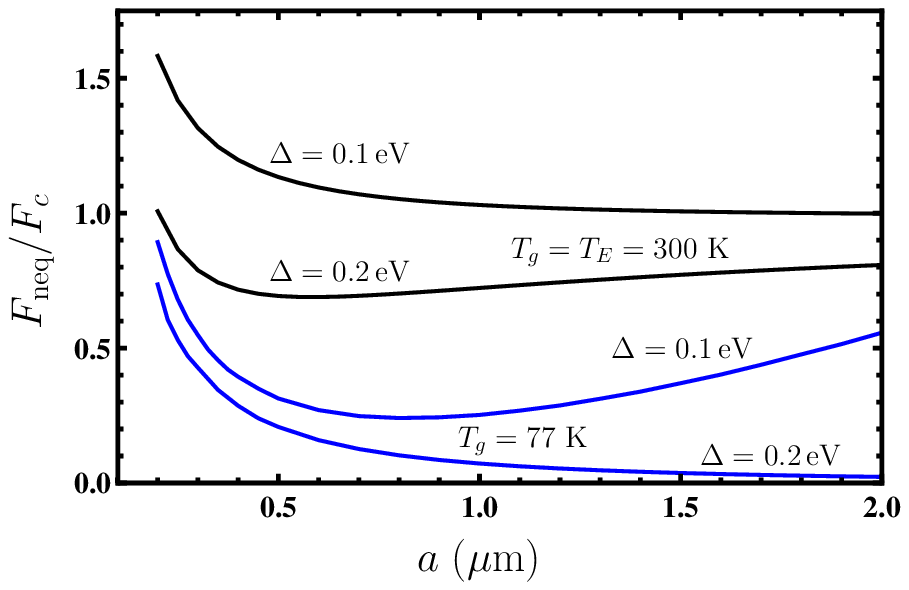}}
\vspace*{-3.7cm}
\caption{\label{fg1} The ratio of nonequilibrium force acting on a nanoparticle
on the source side of cooled to $T_g=77~$K gapped graphene sheets with
$\Delta=0.2$ and 0.1~eV to the classical limit of an equilibrium
at $T_E=300~$K force acting between the same nanoparticle and an ideal
metal plane is shown by the two blue bottom lines as the function of
separation. The two black top lines show the ratio of the equilibrium
force between a nanoparticle and the gapped graphene sheets kept at
$T_E=300~$K to the same classical limit.}
\end{figure}

In Figure \ref{fg1}, the ratio the $F_{neq}/F_c$ is shown as a
function of the nanoparticle-graphene separation by the two bottom
(blue) lines plotted for graphene sheets with the mass-gap parameter
$\Delta$ equal to 0.2 and 0.1 eV kept at temperature $T_g=77$K.
%%%%%%%%%%%%%%%%%%%%%%__Fig._1__%%%%%%%%%%%%%%%%%%
%%%%%%%%%%%%%%%%%%%%%%%%%%%%%%%%%%%%%%%%%%

For comparison purposes, the two top (black) lines in Figure \ref{eq1}
show the ratio $F_{\rm eq}/F_c$ for the same nanoparticles and graphene
sheets computed in the state of thermal equilibrium, i.e., when the
temperatures of graphene and nanoparticles are equal to the environmental
temperature, $T_g=T_E=300~$K. In this case, the quantity $F_{eq}$ is
computed as

\begin{eqnarray}
&&
F_{\rm eq}(a,\Delta,T_E)=-\frac{2k_BT_E\alpha(0)}{c}
\sum_{l=0}^{\infty}{\vphantom{\sum}}^{\prime}
\int\limits_0^{\infty}k\,dke^{-2aq_l(k)}
\label{eq31}\\
&&~~~\times
\left\{\left[2q_l^2(k)c^2-\xi_{E,l}^2\right]
R_{\rm TM}(i\xi_{E,l},k;\Delta,T_E)-\xi_{E,l}^2R_{\rm TE}(i\xi_{E,l},k;\Delta,T_E)\right\}.
\nonumber
\end{eqnarray}

\noindent
Equation (\ref{eq31}) is obtained from (\ref{eq3}) by putting $T_g=T_E$.

As is seen in Figure \ref{fg1}, for a cooled graphene sheet the change
in the value of $\Delta$ makes a lesser impact on $F_{\rm neq}$ than on
$F_{\rm eq}$ at short separations but, on the contrary, makes a greater
impact on $F_{\rm neq}$ than on $F_{\rm eq}$ at large separations. As opposed
to the case of cooled sheet of a pristine graphene \cite{65}, for a
gapped graphene with sufficiently large $\Delta$ the nonequilibrium
dispersion force remains attractive.

\begin{figure}[H]
\vspace*{-7.2cm}
\centerline{\hspace*{-2.7cm}
\includegraphics[width=4.5in]{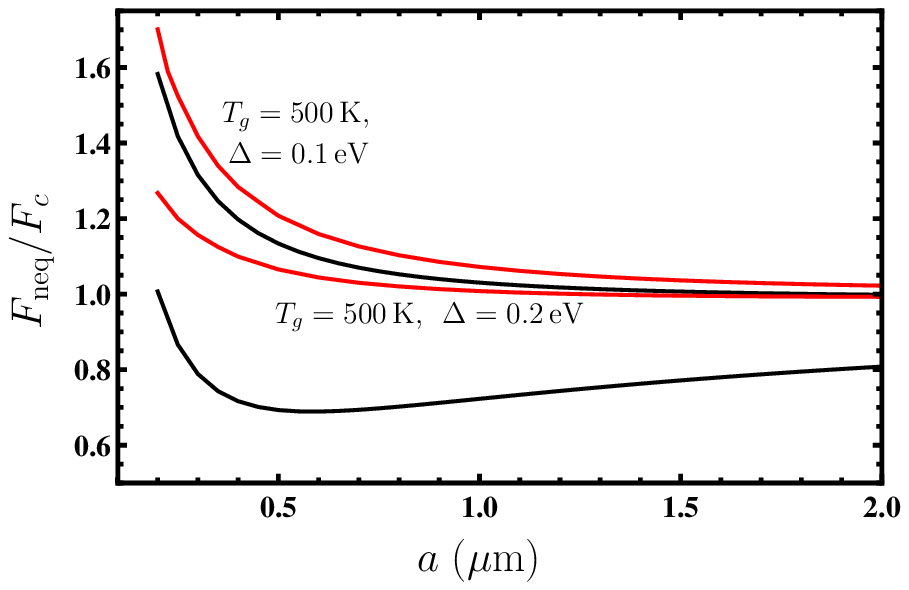}}
\vspace*{-3.7cm}
\caption{\label{fg2} The ratio of nonequilibrium force acting on a nanoparticle
on the source side of heated to $T_g=500~$K gapped graphene sheets with
$\Delta=0.2$ and 0.1~eV to the classical limit of an equilibrium at
$T_E=300~$K force acting between the same nanoparticle and an ideal metal
plane is shown by the two red lines as the functions of separation.
The two black lines reproduced from Figure 1 show the ratio of the
equilibrium force between a nanoparticle and the gapped graphene sheets
kept at $T_E=300~$K to the same classical limit.}
\end{figure}

Now let us admit that the graphene sheet is heated up to $T_g=500~$K,
whereas nanoparticles preserve the environmental temperature $T_E=300$K.
In this case, the computational resluts for the ratio $F_{\rm neq}/F_c$
are shown in Figure \ref{fg2} as the functions of separation by the
two red lines plotted for graphene sheets with the mass-gap parameter
$\Delta$ equal to 0.2 and 0.1~eV. The two black lines, which show the
ratio $F_{\rm eq}/F_c$, are reproduced from Figure \ref{fg1}. As explained
above, they are plotted for graphene sheets with $\Delta=0.2$ and 0.1~eV
in thermal equilibrium with the environment at $T_g=T_E=300~$K.

{}From Figure \ref{fg2}, it is seen that for a heated graphene sheet the
change in the value of $\Delta$ makes a lesser impact on $F_{neq}$ than
on $F_{eq}$ over the entire separation region considered. By comparing
Figures \ref{fg1} and \ref{fg2}, one can conclude that the magnitude of
a nonequilibrium dispersion force acting on a nanoparticle on the source
side of gapped graphene increases with increasing temperature.
%%%%%%%%%%%%%%%%%%%%%%__Fig._2__%%%%%%%%%%%%%%%%%%
%%%%%%%%%%%%%%%%%%%%%%%%%%%%%%%%%%%%%%%%%%

\begin{figure}[H]
\vspace*{-7.2cm}
\centerline{\hspace*{-2.7cm}
\includegraphics[width=4.5in]{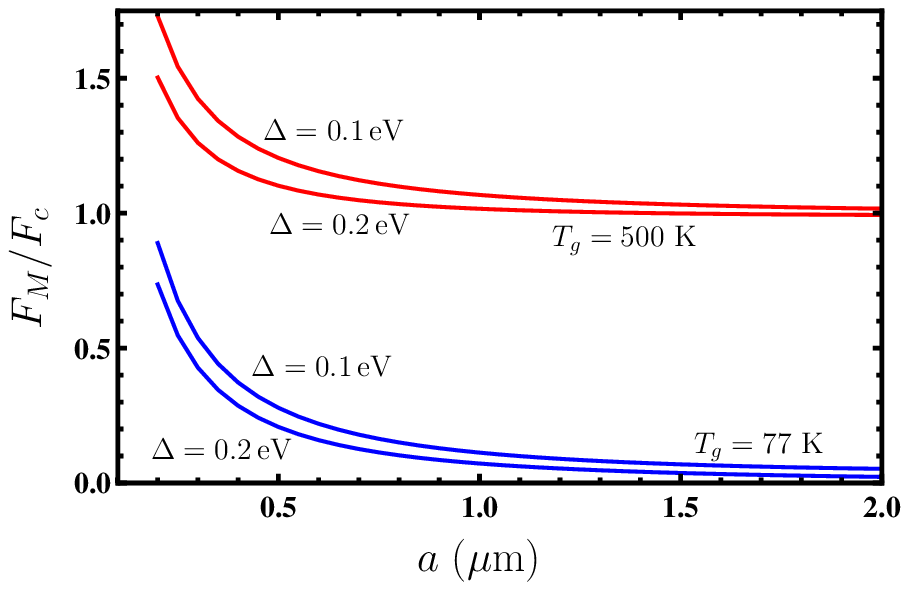}}
\vspace*{-3.7cm}
\caption{\label{fg3} The ratio of the first contribution to the nonequilibrium
force acting on a nanoparticle on the source side of cooled to $T_g=77~$K
and heated to $T_g=500~$K gapped graphene sheets to the classical limit of
an equilibrium at $T_E=300~$K force acting between the same nanoparticle
and an ideal metal plane is shown by the pairs of blue and red lines,
respectively, as the function of separation. In each pair, the lower line
is for a graphene sheet with the mass-gap parameter $\Delta=0.2~$eV and the upper
line is for a graphene sheet with $\Delta=0.1~$eV.}
\end{figure}

Now we investigate the relative role of the first and second contributions
$F_M$ and $F_r$ in (\ref{eq2}), which sum represents the total value of
$F_{neq}$. We begin with $F_M$ computed by Equations (\ref{eq3}) and
(\ref{eq4}) and respective expressions (\ref{eq8}),
(\ref{eq24})--(\ref{eq28}) for the polarization tensor calculated at the
pure imaginary Matsubara frequencies i$\xi_{E,l}$.
Similar to the case of an equilibrium force, $F_M$ is always negative, i.e.,
contributes to the attraction.
The computational
results for the ratio $F_M/F_c$ are shown as the functions of separation
in Figure \ref{fg3} by the two pairs of blue and red lines computed at
the graphene temperature $T_g=77~$K and 500 K, respectively. In each pair,
the lower line is for a graphene sheet with $\Delta=0.2~$eV and the upper
line is for a graphene sheet with $\Delta=0.1~$eV.

As is seen in Figure \ref{fg3}, at both temperatures the contribution
$F_M$ decreases in magnitude with increasing separation.
In the separation region considered,  this decrease
occurs to the relatively small values at $T_g=77~$K and to the classical limit (\ref{eq30})
at $T_g=500$K. In doing so, the sign of $F_M$ remains negative which
corresponds to the attractive force.

The role of the contribution $F_r$ is somewhat different.
The sign of $F_r$ in (\ref{eq6}) is determined by the sign of the quantity $\Theta$ in
(\ref{eq7}) and of the imaginary parts of $R_{\rm TM}$ and $R_{\rm TE}$ defined
in (\ref{eq4}). Thus, $F_r$ can be both negative and positive. The quantity $\Theta$
is negative for $T_g>T_E$ and is positive for $T_g<T_E$. As to the sign of
${\rm Im}\,R_{\rm TM}$ and ${\rm Im}\,R_{\rm TE}$, it depends on the relative
contributions of the frequency regions (\ref{eq9}) and (\ref{eq10}).
In the region (\ref{eq9}),  ${\rm Im}\,R_{\rm TM,TE}$ is positive and in the region
(\ref{eq10}) --- negative.
The computational
results for $F_r$ are obtained by Equations (\ref{eq6}), (\ref{eq4}) and
respective expressions for the polarization tensor at real frequencies
presented in Section 3.

%%%%%%%%%%%%%%%%%%%%%%__Fig._3__%%%%%%%%%%%%%%%%%%
%%%%%%%%%%%%%%%%%%%%%%%%%%%%%%%%%%%%%%%%%%

\begin{figure}[H]
\vspace*{-3.5cm}
\centerline{\hspace*{-1.cm}
\includegraphics[width=7.5in]{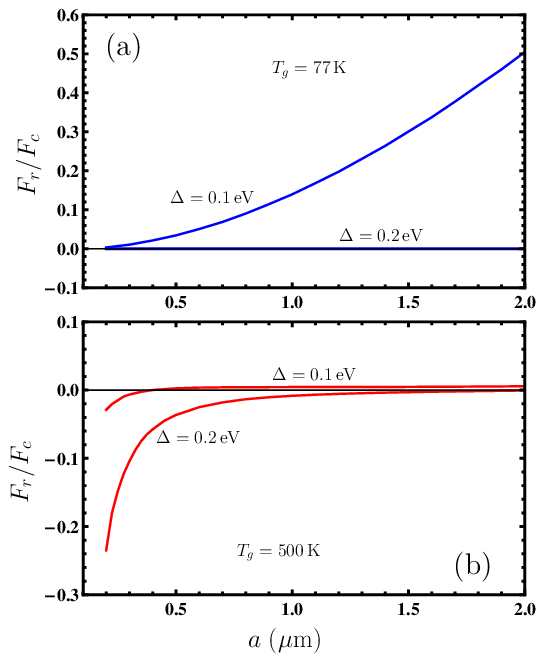}}
\vspace*{-13.cm}
\caption{\label{fg4} The ratio of the second contribution to the nonequilibrium
force acting on a nanoparticle on the source side of (a) cooled to
$T_g=77~$K and (b) heated to $T_g=500~$K gapped graphene sheets to the
classical limit of an equilibrium at $T_E=300~$K force acting between
the same nanoparticle and an ideal metal plane is shown by the pairs of
blue and red lines, respectively, as the function of separation. In each
pair, the lower line is for a graphene sheet with the mass-gap parameter
$\Delta=0.2~$eV and the upper line is for a graphene sheet with
$\Delta=0.1~$eV.}
\end{figure}

In Figure \ref{fg4}(a), we plot the ratio $F_r/F_c$
as the function of separation at the graphene temperature $T_g=77~$K and
in Figure \ref{fg4}(b) --- for $T_g=500~$K (the blue and red pairs of lines,
respectively). In both cases the lower and upper lines are for the graphene
sheets with $\Delta=0.2$ and 0.1~eV, respectively.
{}From Figure \ref{fg4}(a) one can see that for a graphene sheet with
$\Delta=0.2~$eV at $T_g=77~$K the contribution $F_r$ remains negligibly
small at all separations considered, whereas it increases monotonously with
increasing separation for a graphene sheet with $\Delta=0.1~$eV. The sign
of $F_r$ remains negative. Here, the main contribution to $F_r$ given by
the frequency region (\ref{eq10}) is negative leading to $F_r<0$.

In Figure \ref{fg4}(b), the sign of $F_r$ is positive for a graphene sheet
with $\Delta=0.2~$eV and changes from the positive to negative for graphene
with $\Delta=0.1~$eV. This means that, at the separations considered,
the main contribution to $F_r$ for a graphene sheet with $\Delta=0.2~$eV
given by the frequency region (\ref{eq10}) is positive leading to $F_r>0$.
If $\Delta=0.1~$eV, the
relative role of the frequency regions (\ref{eq9}) and (\ref{eq10}) is
different depending on separation. At short distances the dominant region
is (\ref{eq10}) and $F_r>0$, whereas at separations exceeding approximately
$0.5~\mu$m the dominant contribution is given by the region (\ref{eq9})
and $F_r<0$.

By comparing Figure \ref{fg1} with Figures \ref{fg3} and \ref{fg4}(a), it
is seen that at short separations the major contribution to $F_{\rm neq}$ for
a graphene sheet at $T_g=77~$K is given by $F_M$ for both values of the
mass-gap parameter. At large separations, the major contribution to
$F_{\rm neq}$ is given by $F_r$ for graphene with $\Delta=0.1~$eV, whereas the
relatively small values of $F_{\rm neq}$ for graphene with $\Delta=0.2~$eV are
determined by $F_M$.

In a similar way, by comparing Figure \ref{fg2} with Figures \ref{fg3}
and \ref{fg4}(b), we conclude that at short separations the major
contributions to $F_{\rm neq}$ for graphene sheets at $T_g=500~$K with both
values of $\Delta$ are also given by $F_M$. These contributions,
however, are slightly decreased by the impact of $F_r$ which is of the
opposite sign. At large separations the major contribution to $F_{\rm neq}$
is again given by $F_M$ for both values of $\Delta$, but for $\Delta=0.1~$eV
its magnitude is slightly increased at the expense of $F_r$.

At the intermediate separation distances, the value of the nonequilibrium
dispersion force acting on a nanoparticle on the source side of gapped
graphene sheet is determined by the joint action of both contributions
$F_M$ and $F_r$.

%%%%%%%%%%%%%%%%%%%%%%__Fig._4__%%%%%%%%%%%%%%%%%%
%%%%%%%%%%%%%%%%%%%%%%%%%%%%%%%%%%%%%%%%%%

%%%%%%%%%%%%%%%%%%%%%%%%%%%%%%%%%%%%%%%%%%%%%%%%%
\section{Discussion}

In this article, we have investigated the dispersion (Casimir-Polder)
force acting on a nanoparticle on the source side of a gapped graphene
sheet in the nonequilibrium situations when the graphene temperature
is not equal to the nanoparticle temperature coinciding with the
temperature of the enovironment. Both cases when the graphene
temperature is lower and higher than that of the environment were
considered.

It was shown that the nonzero value of the mass-gap parameter results
in new properties of the nonequilibrium dispersion force as compared to
the case of thermal equilibrium. Specifically, for a cooled graphene
sheet, the variation of the mass-gap parameter makes a lesser and
greater impact on the nonequilibrium force than on the equilibrium one
at short and large separations, respectively. For a heated graphene
sheet, the variation of the mass-gap parameter results in a lesser
impact on the nonequilibrium force than on the equilibrium one at
all separations considered from 200 nm to 2$\mu$m. As opposed to the
case of a pristine graphene, for a gapped graphene sheet, the
nonequilibrium dispersion force preserves an attractive character
at all separations considered.

We emphasize that the above results were obtained using the
dielectric response of graphene expressed via the polarization tensor.
The latter quantity was found in the framework of the Dirac model on
the solid foundation of quantum field theory with no recourse to any
phenomenological methods. Thus, in the application region of the
Dirac model discussed in Section 3, these results possess a highest
degree of reliability. In fact graphene and other 2D materials, such
as silicene, stanene, germanene etc. \cite{87,88,89,90,91,92}, are unique in
that some of their properties can be investigated basing on the most
fundamental physical principles. The nonequilibrium dispersion force
acting on nanoparticles on the source side of gapped graphene
considered above presents one more example of this kind.

%%%%%%%%%%%%%%%%%%%%%%%%%%%%%%%%%%%%%%%%%%
\section{Conclusions}

To conclude, the above results give the possibility to control the
nonequilibrium dispersion interaction between nanoparticles and a
graphene sheet by varying the mass-gap parameter of this sheet and
its temperature. The need for such a control is apparent when taken
into account that both nanoparticles of different kinds and graphene
are already widely used in various micro- and nanodevices, including
the field-effect transistors, integrated nanoparticle-biomolecule
systems, electrochemical sensors and biosensors etc.
\cite{6,7,16,17,37,38,39,40,41,42,43,44}. The theoretical methods
used in the design of these micro- and nanodevices are often based
on the phenomenology and computer simulation, rather than on the
fundamental physical principles. It is hoped that an employment
of the methods of fundamental physics will further accelerate the
progress in this rapidly developing field of applied science.

In the future, it would be interesting to extend the obtained results
to graphene sheets deposited on substrates made of metallic and
dielectric materials and to consider the case of doped graphene
characterized by some nonzero chemical potential. This will provide
further possibilities to control the nonequilibrium dispersion
interaction in micro- and nanodevices incorporating nanoparticles
and graphene sheets for their functionality.

%%%%%%%%%%%%%%%%%%%%%%%%%%%%%%%%%%%%%%%%%%
\vspace{6pt}

%%%%%%%%%%%%%%%%%%%%%%%%%%%%%%%%%%%%%%%%%%

\funding{The work of O.Yu.T., was supported by the Russian Science Foundation
under Grant No. 21-72-20029. G.L.K. was partially funded by the
Ministry of Science and Higher Education of Russian Federation
("The World-Class Research Center: Advanced Digital Technologies,"
contract No. 075-15-2022-311 dated April 20, 2022). The research
of V.M.M. was partially carried out in accordance with the Strategic
Academic Leadership Program "Priority 2030" of the Kazan Federal
University. }

%%%%%%%%%%%%%%%%%%%%%%%%%%%%%%%%%%%%%%%%%%
%%%%%%%%%%%%%%%%%%%%%%%%%%%%%%%%%%%%%%%%%%
\end{paracol}

\reftitle{References}
%=====================================

%%%%%%%%%%%%%%%%%%%%%%%%%%%%%%%%%%%%%%%%%%
\end{document}